\documentclass[smallextended]{svjour3}
\usepackage[english]{babel}
\usepackage{graphicx}
\begin{document}
\title{Scaling properties of the number of random sequential adsorption iterations needed to generate saturated random packing}
\author{Micha\l{} Cie\'sla}
\institute{M. Smoluchowski Institute of Physics, Jagiellonian University, \L{}ojasiewicza 11, 30-348 Cracow, Poland}
%
\date{Received: date / Accepted: date}
%
\maketitle
\begin{abstract}
The properties of the number of iterations in random sequential adsorption protocol needed to generate finite saturated random packing of spherically symmetric shapes were studied. Numerical results obtained for one, two, and three dimensional packings were supported by analytical calculations valid for any dimension $d$. It has been shown that the number of iterations needed to generate finite saturated packing is subject to Pareto distribution with exponent $-1-1/d$ and the median of this distribution scales with packing size according to the power-law characterized by exponent $d$. Obtained results can be used in designing effective RSA simulations.
\end{abstract}
%
%
\section{Introduction}
Random sequential adsorption (RSA) is one of numerical protocols that allow to generate random packing of any objects~\cite{feder1980,evans1993}. It is based on the iteration of two simple steps:
\begin{description}
\item[-] a virtual particle is created. Its position and orientation within a packing is selected randomly.
\item[-] the virtual particle is tested for overlaps with any of the other particles in the packing. If no overlap is found, it is added to the packing and holds its position and orientation until the end of simulations. Otherwise, the virtual particle is removed from the packing and abandoned.
\end{description}
A random packing is saturated when there is no possibility to place another particle in it.
The name of RSA is related to irreversible adsorption processes where particles are randomly placed on a surface or interface \cite{dabrowski2001,adamczyk2012,kujda2016} -- one iteration of RSA algorithm corresponds to one adsorption attempt of a single particle. However, RSA applications are much wider including soft matter~\cite{torquato2010}, telecommunication~\cite{hastings2005}, information theory~\cite{coffman1998} and mathematics~\cite{zong2014}.

The main problem of RSA is its efficiency when packing is almost saturated. For such packings the probability of placing another particle is very small. Therefore, a large number of iterations is needed to add a particle and packing's growth becomes very slow. According to Feder's law, density of particles in an infinite packing changes with the number of adsorption attempts according to the following relation~\cite{pomeau1980,swendsen1981,hinrichsen1986,cadilhe2007}:
\begin{equation}
\rho(t) = \rho - A t^{-1/d}.
\label{eq:fl}
\end{equation}
Here, $\rho$~is the density of particles in a saturated packing, $\rho(t)$~is the density of particles in a packing after time $t$, $A$~is a positive constant and $d$ equals to packing dimension in case of spherically symmetric shapes (line segments in one dimension, disks in two dimensions, spheres in three dimensions, etc.). 

The time $t$ equals to the number of attempts of adding a particle per unit length for one dimensional packings,  area for two dimensional packings, volume for three dimensional packings, etc. 

The relation (\ref{eq:fl}) was tested numerically to be valid for large enough, but finite packings~\cite{feder1980,torquato2006,ciesla2012,ciesla2013}. It is commonly used to estimate the number of particles at jamming, however it does not give any hints related to the number of RSA iterations needed to saturate a packing.

Recently, Zhang et al. have improved RSA algorithm, and showed that it is possible to generate saturated random packings of spherically symmetric particles in a reasonable simulation time \cite{zhang2013}. The algorithm is based on tracking the area where placing subsequent particles is possible. The idea comes from earlier works concerning RSA on discrete lattices~\cite{nord1991} and deposition of oriented squares on a continuous surface \cite{brosilow1991}. When particle is added to the packing this available area decreases. When it vanishes completely the packing is saturated. Because a single sampling of a particle centre from the available area of size $s$ is equivalent to $S/s$ samplings from the whole packing, this method makes it possible to determine number of RSA iterations in original protocol needed to saturate finite packing. 

The aim of this paper is to analyse properties of this number as a random variable. In particular, it is interesting to investigate how it depends on packing size, since a recent study suggests that its median scales according to the power-law with an exponent equal to packing dimension \cite{ciesla2016}.
\section{Details of numerical simulations}
Saturated random packing were generated using method described in detail in~\cite{zhang2013}. Line segments, disks and spheres, were packed onto a large line segment, square, and cube, respectively. In all these cases, periodic boundary conditions were used. The size (length, area or volume depending on a packing dimension) of packings varied from $10^4$ to $4 \times 10^7$ and a single shape had a unit volume. For each particular set-up at least $100$ of independent packings were created. For each of the saturated packings, the value of $t$ was recorded at the moment when the last particle was added.
\section{Results and discussion}
\subsection{Distribution of simulation time}
\begin{figure}[htb]
\vspace{0.5in}
\centerline{%
\includegraphics[width=0.7\columnwidth]{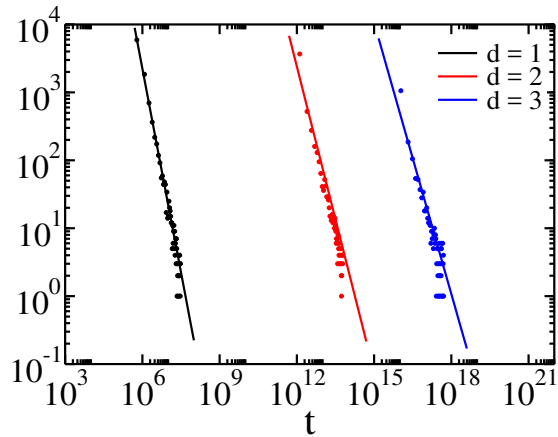}
}
\caption{Histograms of RSA iterations per unit area needed to saturate $d$-dimensional packings.
Dots represent numerical data obtained from numerically generated $10^4$ packings of size $S=10^6$ for $d=1$ and $d=2$, and $S=10^5$ for $d=3$. Lines correspond to power fits: $2.23 \times 10^{15} \cdot t^{-2}$, $2.41 \times 10^{21} \cdot t^{-1.5}$ and $1.06 \times 10^{24} \cdot t^{-1.333}$ for $d=1$, $2$, and $3$, respectively.}
\label{fig:histogram}
\end{figure}

\begin{figure}[htb]
\vspace{0.5in}
\centerline{%
\includegraphics[width=0.4\columnwidth]{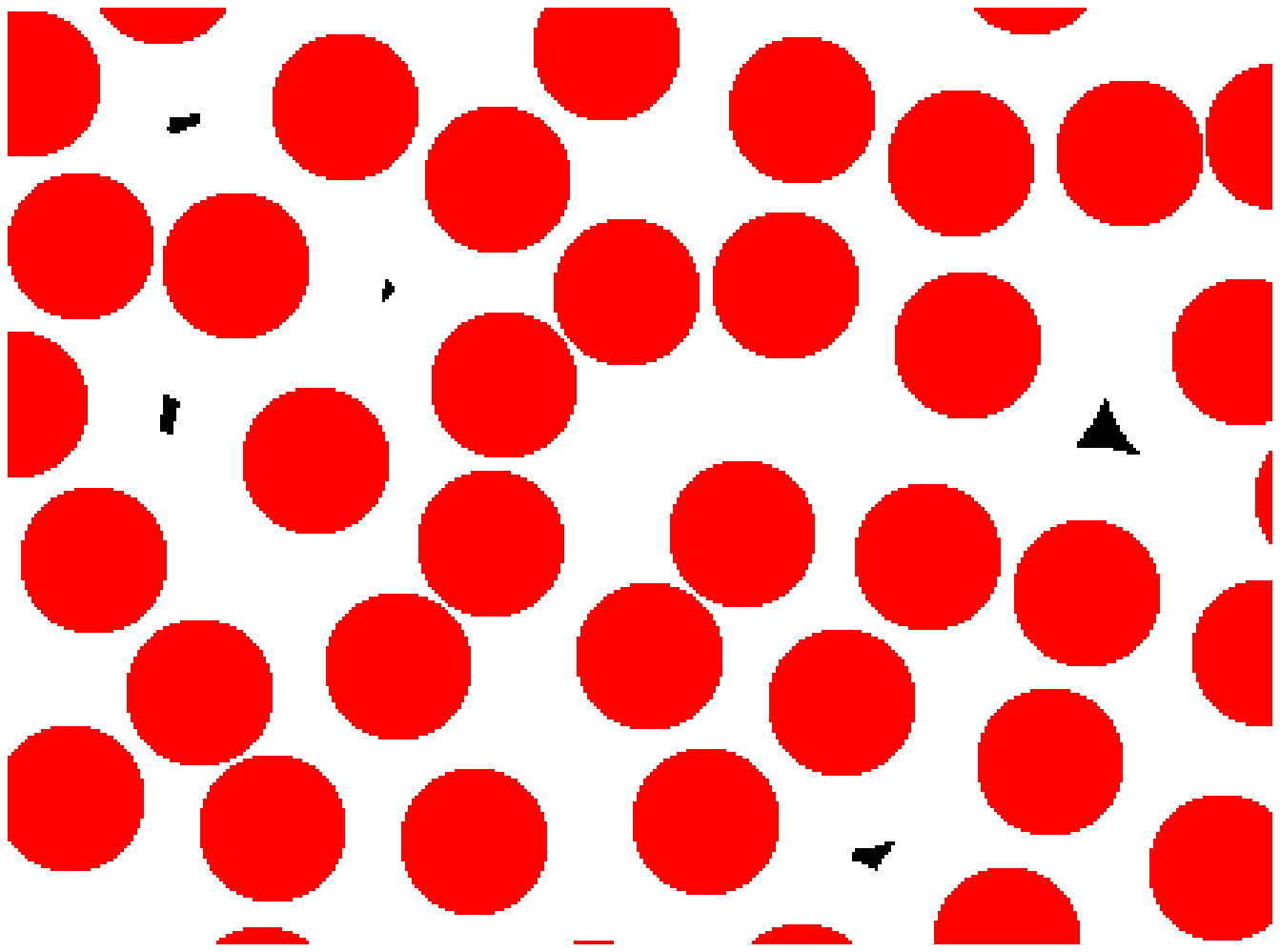}
\hspace{0.05\columnwidth}
\includegraphics[width=0.3\columnwidth]{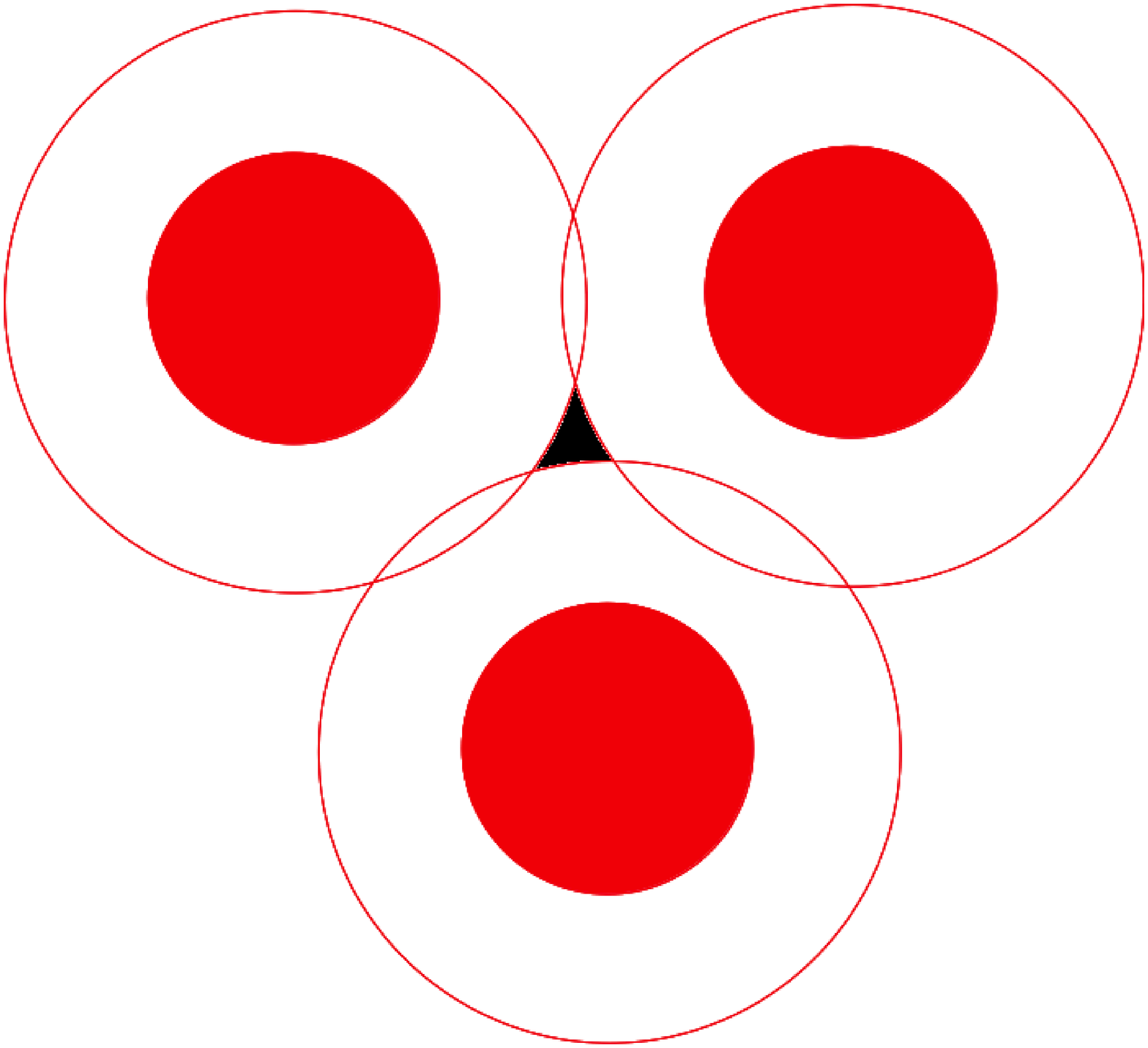}
}
\caption{Fragment of a non-saturated two-dimensional packing of disks with regions (the black triangle-like areas) where centres of subsequent disks can be placed. On the right the magnification of one such region. The diameter of red circles is twice as large as the diameter of red disks.}
\label{fig:region}
\end{figure}

Example of distributions of time needed to get saturated packings of line segments, disks, and spheres obtained from RSA simulations are shown in Fig.\ref{fig:histogram}.
To study the time after which simulation ends lets focus on the probability of placing a particle in the packing, as the inverse of this probability is proportional to the expected number of iterations needed to find a large enough place for that particle. This probability has already been studied by Pomeau \cite{pomeau1980} and Swendsen \cite{swendsen1981}. For sufficiently long times, in a non-saturated packing, there are separate regions where subsequent particles can be placed (see Fig. \ref{fig:region})
Swendsen assumed that independently of packing dimension, the linear size of such regions is uniformly distributed \cite{swendsen1981}, which has been recently confirmed in numerical experiments for two dimensional random packings \cite{ciesla2016}. This assumption leads to the following distribution of regions sizes:
\begin{equation}
p_\mathrm{s}(x) = \frac{x^{-\frac{d - 1}{d}}}{d \, {s_{0}}^\frac{1}{d}}
\label{eq:ps}
\end{equation}
where $s$ is a random variable denoting the volume of a region, $s_0$ is the volume of the largest region, and 
\begin{equation}
  p_\mathrm{s}(x) \equiv \lim_{dx\to 0} \frac{Prob(x < s < x+dx)}{dx}
\end{equation}
To saturate a packing all such regions should be filled. Probability of placing a centre of a particle inside a region is proportional to its area. Because these regions are well separated, placing a subsequent disk in one region does not affect the probability of filling any other region. This means that, statistically, smaller regions will be filled later than larger ones; thus, the last region to be filled should be the smallest one. Therefore, the distribution of iterations needed to generate saturated packing is directly related to the distribution of the sizes of the smallest region. As the cumulative distribution function of region area size is
\begin{equation}
F_\mathrm{s}(x) \equiv Prob(s<x) = \int_0^x p_\mathrm{s}(t) dt = \left( \frac{x}{s_{0}} \right)^{\frac{1}{d}}.
\end{equation}
The probability that a random variable $s_\mathrm{min}$, being the minimum of $n$ independent random variables $s$ is smaller than a given value x is:
\begin{eqnarray}
Prob(s_\mathrm{min}<x) & = & 1 - \underbrace{Prob(s>x)\cdot Prob(s>x) \cdot \dots \cdot Prob(s>x)}_{n \,\,\mathrm{ times}} \nonumber \\
 & = & 1 - \left[ 1 - Prob(s<x) \right]^n. 
\end{eqnarray}
Thus, the cumulative distribution function of the $s_\mathrm{min}$, is equal to:
\begin{equation}
F_\mathrm{s_\mathrm{min}}(x) = 1 - \left[ 1 - F_\mathrm{s}(x) \right]^n = 1 - \left[ 1 - \left( \frac{x}{s_{0}} \right)^{\frac{1}{d}} \right]^n,
\label{eq:fsmin}
\end{equation}
and the probability distribution function of $s_\mathrm{min}$ is
\begin{equation}
p_\mathrm{s_\mathrm{min}}(x) = \frac{n}{d {s_0}^{\frac{1}{d}}} \, x^{\frac{1}{d}-1} \left[ 1 - \left(\frac{x}{s_0}\right)^\frac{1}{d}\right]^{n-1}.
\label{eq:psmin}
\end{equation}
Note that asymptotically for $x \ll s_0$, the expression in square brackets is practically equal to $1$, thus, $p_\mathrm{s_\mathrm{min}}(x) \sim x^{1/d - 1}$.
As noted previously, the number of iterations $t_\mathrm{max}$ needed to create saturated packing is inversely proportional to $s_\mathrm{min}$: $t_\mathrm{max} \sim 1 / s_\mathrm{min}$. Thus, taking into account the above relation, asymptotically
\begin{equation}
p_\mathrm{t_\mathrm{max}}(x) \sim x^{-1-\frac{1}{d}},
\label{eq:ptmax}
\end{equation}
which is in a quite good agreement with numerical experiments (see Fig.\ref{fig:histogram}).
\subsection{Median of simulation time}
\begin{figure}[htb]
\vspace{0.5in}
\centerline{%
\includegraphics[width=0.7\columnwidth]{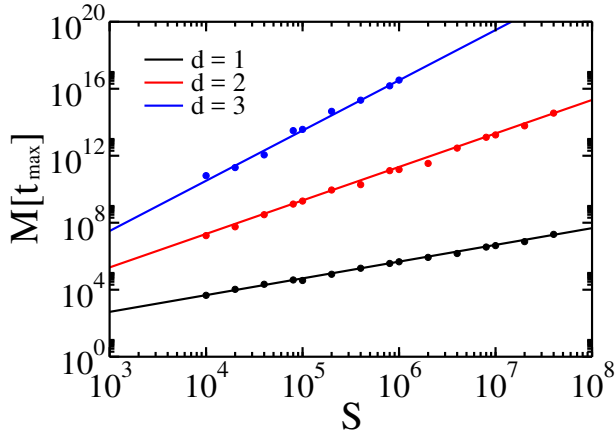}
}
\caption{The dependence of the median of dimensionless time needed to obtain saturated packing on packing size for segments ($d=1$), disks ($d=2$), and spheres ($d=3$). Dots represent data obtained from numerical analysis of $10^2$ packings of size $S \in [10^4, 4 \times 10^8]$. Lines corresponds to fits: $0.474587 \cdot S$, $0.214991 \cdot S^2$ and $0.0320608 \cdot S^3$ for $d=1$, $2$, and $3$, respectively.}
\label{fig:median}
\end{figure}

To explain recently observed scaling of the median of $t_\mathrm{max}$ with packing size \cite{ciesla2016}, it is necessary to move back to Eq.(\ref{eq:fsmin}) and solve the equation $F_\mathrm{s_\mathrm{min}}(x) = 1/2$, which gives the median of $s_\mathrm{min}$.
\begin{equation}
M[s_\mathrm{min}] = s_0 \left[1 - \left(\frac{1}{2}\right)^{\frac{1}{n}} \right]^d
\end{equation}
Again, as $t_\mathrm{max} \sim 1 / s_\mathrm{min}$, which is monotonic, then $M[t_\mathrm{max}] \sim 1 / M[s_\mathrm{min}]$, thus
\begin{equation}
M[t_\mathrm{max}] \sim \left[1 - \left(\frac{1}{2}\right)^{\frac{1}{n}} \right]^{-d}
\end{equation}
Number of regions $n$ that have to be filled scales linearly with packing size $S$. Moreover, the expression in square brackets may be expanded in the series $[1 - (1/2)^{1/n}] = \ln 2 / n - 1/2 (\ln 2 / n)^2 + o(1/n^2)$. Thus, asymptotically for large S (and small $1/n$):
\begin{equation}
M[t_\mathrm{max}] \sim \left[ \frac{\ln 2}{S} \right]^{-d} \sim S^d
\label{eq:Mtmax}
\end{equation}
Numerical confirmation of this result is shown in Fig.\ref{fig:median}.
It is rather obvious that generating larger packings needs more computational time. But, the plot also shows that RSA in higher dimensions may be extremely time consuming as the time needed to generate a packing of the same size grows with dimension, as well as it scales with packing size with higher exponent. Moreover, packing fraction decreases with the growth of packing dimension \cite{zhang2013,torquato2006}, which additionally spoils statistics in numerical simulations. Additionally, it is worth noting that Feder's law (\ref{eq:fl}) seems to be valid also for fractional packing's dimensions \cite{ciesla2012,ciesla2013}. As its derivation bases on the same assumption as made here, namely (\ref{eq:ps}), presented results should be valid also for fractional $d$'s. At last, the relation (\ref{eq:fl}) works also for anisotropic shapes \cite{viot1992,ciesla2016shapes}, with parameter $d$ denoting a number of degrees of freedom of adsorbed particle instead of packing dimension \cite{ciesla2013}. Therefore, the question arises if relations (\ref{eq:ptmax}) and (\ref{eq:Mtmax}) are valid for anisotropic particles. But to answer it, the algorithm that generates saturated random packing in reasonable computational time is needed.
\section{Conclusions}
The number of RSA iterations needed to generate finite saturated random packing of spherically symmetric objects seems to subject to Pareto distribution with an exponent $-(1+1/d)$ where $d$ is packing dimension. The median of this distribution scales with packing size according to the power-law characterized by exponent $d$. Obtained results can be used to optimize and control time complexity of RSA simulations. In particular, together with results described in \cite{ciesla2016}, they allow to design an RSA simulation that gives possibly the most accurate results within existing time limits.
\section*{Acknowledgement}
I would like to thank Grzegorz Paj\c{a}k for inspiring and fruitful discussions and Robert~M.~Ziff for comments on the paper. Part of numerical simulations was carried out with the support of the Interdisciplinary Centre for Mathematical and Computational Modeling (ICM) at University of Warsaw under grant no.\ G-27-8.
\bibliography{main.bib}
\end{document}